\documentclass[aps,prd,twocolumns,eqsecnum,preprintnumbers,showpacs,amsmath,amssymb]
{revtex4}

\usepackage{graphicx}
\usepackage{dcolumn}
\usepackage{bm}

\begin{document}

\title{THE MASS GAP AND GLUON CONFINEMENT }

\author{V. Gogohia}
\email[]{gogohia@rmki.kfki.hu}

\affiliation{HAS, CRIP, RMKI, Depart. Theor. Phys., Budapest 114,
P.O.B. 49, H-1525, Hungary}

\date{\today}
\begin{abstract}
In our previous publications \cite{1,2,3} it has been proven that
the general iteration solution of the Schwinger-Dyson equation for
the full gluon propagator (i.e., when the skeleton loop integrals,
contributing into the gluon self-energy, have to be iterated,
which means that no any truncations/approximations have been made)
can be algebraically (i.e., exactly) decomposed as the sum of the
two principally different terms. The first term is the Laurent
expansion in integer powers of severe (i.e., more singular than
$1/ q^2$) infrared singularities accompanied by the corresponding
powers of the mass gap and multiplied by the corresponding
residues. The standard second term is always as much singular as
$1/q^2$ and otherwise remaining undetermined. Here it is
explicitly shown that the infrared renormalization of the mass gap
only is needed to render theory free of all severe infrared
singularities in the gluon sector. Moreover, this leads to the
gluon confinement criterion in a gauge-invariant way. As a result
of the infrared renormalization of the mass gap in the initial
Laurent expansion, that is dimensionally regularized, the simplest
severe infrared singularity $(q^2)^{-2}$ survives only. It is
multiplied by the mass gap squared, which is the scale responsible
for the large scale structure of the true QCD vacuum. The
$\delta$-type regularization of the simplest severe infrared
singularity (and its generalization for the multi-loop skeleton
integrals) is provided by the dimensional regularization method
correctly implemented into the theory of distributions. This makes
it possible to formulate exactly and explicitly the full gluon
propagator (up to its unimportant perturbative part).
\end{abstract}

\pacs{ 11.15.Tk, 12.38.Lg}

\keywords{}

\maketitle

\section{Introduction}

The propagation of gluons is one of the main dynamical effects in
the true QCD vacuum. The gluon Green's function is (Euclidean
signature here and everywhere below)

\begin{equation}
D_{\mu\nu}(q) = i \left\{ T_{\mu\nu}(q)d(q^2, \xi) + \xi
L_{\mu\nu}(q) \right\} {1 \over q^2 },
\end{equation}
where $\xi$  is the gauge fixing parameter ($\xi = 0$ - Landau
gauge and  $\xi = 1$ - Feynman gauge) and
$T_{\mu\nu}(q)=\delta_{\mu\nu}-q_{\mu} q_{\nu} / q^2 =
\delta_{\mu\nu } - L_{\mu\nu}(q)$. Evidently, $T_{\mu\nu}(q)$ is
the transverse ("physical") component of the full gluon
propagator, while $L_{\mu\nu}(q)$ is its longitudinal (unphysical)
one. The free gluon propagator is obtained by setting simply the
full gluon form factor $d(q^2, \xi)=1$ in Eq. (1.1), i.e.,

\begin{equation}
D^0_{\mu\nu}(q) = i \left\{ T_{\mu\nu}(q) + \xi L_{\mu\nu}(q)
\right\} {1 \over q^2}.
\end{equation}
The dynamical equation of motion for the full gluon propagator
(1.1) is known as the gluon Schwinger-Dyson (SD) one (see below
Sect. 3), and its solutions reflect the quantum-dynamical
structure of the true QCD ground state.

In our previous works \cite{1,2,3} we have investigated the
general iteration solution of the above-mentioned gluon SD
equation for the full gluon propagator (1.1). No any
trancations/approximations have been made since the corresponding
skeleton loop integrals, contributing into the gluon self-energy,
have been iterated. It has been exactly proven that the full gluon
propagator is to be algebraically decomposed into the two
principally different terms as follows:

\begin{equation}
D_{\mu\nu}(q) = D^{INP}_{\mu\nu}(q)+ O_{\mu\nu}(1/q^2),
\end{equation}
where

\begin{equation}
D^{INP}_{\mu\nu}(q, \Delta^2)= i T_{\mu\nu}(q) {\Delta^2 \over
(q^2)^2} \sum_{k=0}^{\infty}(\Delta^2 /q^2)^k \phi_k(\lambda, \nu,
\xi, g^2),
\end{equation}
and $O_{\mu\nu}(1/q^2)$ denotes the terms which are the terms of
the order $1/q^2$ at small $q^2$ (see also section IV below). Here
the superscript "INP" stands for the intrinsically nonperturbative
(NP) part of the full gluon propagator. It is nothing but the
Laurent expansion in integer powers of severe (i.e., more singular
than $1/q^2$) infrared (IR) singularities accompanied by the
corresponding powers of the mass gap $\Delta^2$ and multiplied by
the corresponding residues (which are dimensionless, of course).
They are the sum of all iterations, namely

\begin{equation}
\phi_k(\lambda, \nu, \xi, g^2) = \sum_{m=0}^{\infty} \phi_{k,m}
(\lambda, \nu, \xi, g^2),
\end{equation}
and this clearly shows that an infinite number of iterations (each
iteration) of the skeleton loop integrals invokes each severe IR
singularity and hence the mass gap in the full gluon propagator.
Evidently, the mass gap $\Delta^2$ determines the deviation of the
full gluon propagator from the free one in the deep IR limit ($q^2
\rightarrow 0$). It is worth emphasizing that it has not been
introduced by hand. It is hidden in the skeleton loop
contributions into the gluon self-energy. It explicitly shows up
in the terms dominating the IR structure of the full gluon
propagator (the first term in Eq. (1.3)). The appropriate
regularization procedure have been applied to make the existence
of the mass gap perfectly clear \cite{1}. Let us also recall that
here $\lambda$ and $\nu$ are the dimensionless UV cut-off and the
renormalization point, respectively, while $\xi$ and $g^2$ are the
gauge fixing parameter and the dimensionless coupling constant
squared, respectively. As emphasized in Ref. \cite{1} is all that
matters within our approach is the dependence of the residues on
their arguments and not their concrete values (see below as well).
The perturbative (PT) part of the full gluon propagator
$O_{\mu\nu}(1/q^2)$ remains undetermined, which, however, is not
important as well within our approach (for more detail discussion
of these results see our previous papers \cite{1,2,3}).

The main purpose of this paper is to complete the previous work by
investigating the IR renormalization properties of the theory, in
particular of the mass gap. We will show that the gluon
confinement criterion is the gauge-invariant IR renormalization of
the mass gap only effect within our approach. Moreover, we will
establish uniquely and exactly the INP part of the full gluon
propagator by correctly implementing the dimensional
regularization (DR) method \cite{4} into the theory of
distributions \cite{5}.

\section{IR renormalization of the gluon SD equation}

The power-type severe (or equivalently NP) IR singularities
represent a rather broad and important class of functions with
algebraic singularities. They regularization should be done within
the theory of distributions \cite{5}, complemented by the DR
method \cite{4}. The crucial observation is that the
regularization of these singularities does not depend on their
powers \cite{5,6}, namely

\begin{equation}
(q^2)^{- 2 - k } = { 1 \over \epsilon} \Bigr[
a(k)[\delta^4(q)]^{(k)} + O(\epsilon) \Bigl], \quad \epsilon
\rightarrow 0^+,
\end{equation}
where $a(k)$ is a finite constant depending only on $k$ and
$[\delta^4(q)]^{(k)}$ represents the $k$th derivative of the
$\delta$ function. Here $\epsilon$ is the IR regularization
parameter, introduced within the DR method \cite{4,6}, and which
should go to zero at the end of the computations. We point out
that after introducing this expansion everywhere one can fix the
number of dimensions to four without any further problems. The
important observation is that each NP IR singularity scales as $1/
\epsilon$ as $\epsilon$ goes to zero, and the difference between
them appears only in the residues. Just this plays a crucial role
in the IR renormalization of the theory within our approach (see
below). This regularization expansion takes place only in
four-dimensional QCD with Euclidean signature. In other dimensions
and signature it is more complicated \cite{5,6}.

In the presence of such severe IR singularities all the quantities
which appear in the gluon SD equation should depend in principle
on $\epsilon$. Thus the general IR multiplicative renormalization
(IRMR) program is needed in order to express all the
above-mentioned quantities in terms of their IR renormalized
versions. Fortunately, the gluon SD equation (see below) does not
contain unknown scattering amplitudes, which usually are
determined by an infinite series of the multi-loop skeleton
diagrams. It is a closed system in the sense that there is a
dependence only on the pure gluon vertices, quark- and ghost-gluon
vertices and on the corresponding propagators \cite{6,7,8}. Its IR
renormalization can be carried out on general ground. Symbolically
(however, this is sufficient to perform the IRMR program) it can
be written down as follows \cite{2,6}:

\begin{eqnarray}
D(q) = D^0(q) &-& D^0(q)T_q(q) D(q) - D^0(q)T_{gh}(q)D(q) + D^0(q)
{1 \over 2} T_t D(q) \nonumber\\
&+& D^0(q){1 \over 2} T_1(q)D(q)
+ D^0(q){1 \over 2} T_2(q)D(q) + D^0(q){1 \over 6} T_2'(q)D(q),
\end{eqnarray}
where all the skeleton loop integrals are shown explicitly below

\begin{equation}
T_q(q) = - g^2 \int {i d^4 p \over (2 \pi)^4} Tr [\gamma_{\nu}
S(p-q) \Gamma_{\mu}(p-q, q)S(p)],
\end{equation}

\begin{equation}
T_{gh}(q) =  g^2 \int {i d^4 k \over (2 \pi)^4} k_{\nu} G(k)
G(k-q)G_{\mu}(k-q, q),
\end{equation}

\begin{equation}
T_t = g^2 \int {i d^4 q_1 \over (2 \pi)^4} T^0_4 D(q_1),
\end{equation}

\begin{equation}
T_1(q) =  g^2 \int {i d^4 q_1 \over (2 \pi)^4} T^0_3 (q, -q_1,
q_1-q) T_3 (-q, q_1, q -q_1) D(q_1) D(q -q_1),
\end{equation}

\begin{equation}
T_2(q) =  g^4 \int {i d^4 q_1 \over (2 \pi)^4} \int {i d^n q_2
\over (2 \pi)^4} T^0_4 T_3 (-q_2, q_3, q_2 -q_3) T_3(-q, q_1,
q_3-q_2) D(q_1) D(-q_2)D(q_3) D(q_3 -q_2),
\end{equation}

\begin{equation}
T_2'(q) =  g^4 \int {i d^4 q_1 \over (2 \pi)^4} \int {i d^4 q_2
\over (2 \pi)^4} T^0_4 T_4 (-q, q_1, -q_2, q_3) D(q_1)
D(-q_2)D(q_3),
\end{equation}
and in the last two skeleton loop integrals $q-q_1 +q_2-q_3=0$ as
usual. It is instructive to complete the IRMR program for the full
gluon SD equation (i.e., including the quark and ghost degrees of
freedom) and not only for its Yang-Mills (YM) part.

\subsection{Multiplicative Renormalizability}

The next step is to introduce the IR renormalized quantities. As
noted above, in the presence of such severe IR singularities
(2.1), all the quantities should in principle depend on $\epsilon$
as well, i.e., they become IR regularized. So, one has to put

\begin{eqnarray}
g^2 &=& X(\epsilon) \bar g^2, \quad G(k) = \tilde{Z}_2(\epsilon)
\bar G(k), \quad S(p) = Z_2(\epsilon) \bar S(p), \nonumber\\
G_{\mu}(k, q) &=& \tilde{Z}_1(\epsilon) \bar G_{\mu}(k,q), \quad
\Gamma_{\mu}(p,q) = Z_1^{-1}(\epsilon) \bar \Gamma_{\mu}(p,q),
\nonumber\\
D(q) &=& Z_D(\epsilon) \bar D(q), \nonumber\\
T_3(q,q_1) &=& Z_3(\epsilon) \bar T_3(q,q_1), \nonumber\\
T_4(q,q_1,q_2) &=& Z_4(\epsilon) \bar T_4(q,q_1,q_2).
\end{eqnarray}
In all these relations the quantities with bar are, by definition,
IR renormalized, i.e., they exist as $\epsilon$ goes to zero. In
both quantities, the IR regularized (without bar) and the IR
renormalized (with bar), the dependence on $\epsilon$ is assumed
but not shown explicitly, for simplicity. In the corresponding
IRMR constants this dependence is not omitted in order to
distinguish them clearly from the corresponding UVMR constants.
Since we are interested in the IR renormalization of the SD
equation for the full gluon propagator, it is convenient not to
distinguish between the IR renormalization of its INP and PT parts
at this stage.

Substituting these relations into the gluon SD equation (2.2), and
on account of the explicit expressions for the corresponding
skeleton loop integrals given in Eqs. (2.3)-(2.8), one obtains

\begin{eqnarray}
\bar N_1 \bar D(q) = D^0(q) &-& \bar N_6
D^0(q)\bar T_q(q) \bar D(q) - \bar N_7 D^0(q)\bar T_{gh}(q)\bar
D(q) + \bar N_2 D^0(q) {1
\over 2} \bar T_t \bar D(q) \nonumber\\
&+& \bar N_3 D^0(q){1 \over 2} \bar T_1(q)\bar D(q) + \bar N_4
D^0(q){1 \over 2} \bar T_2(q) \bar D(q) + \bar N_5 D^0(q){1 \over
6} \bar T_2'(q),  \nonumber\\
\end{eqnarray}
if and only if the so-called following IR convergence conditions
hold:

\begin{eqnarray}
Z_D(\epsilon) &=& \bar N_1(\epsilon), \quad X(\epsilon) Z_D^2(\epsilon)=
\bar N_2(\epsilon), \nonumber\\
X(\epsilon) Z_D^3(\epsilon) Z_3(\epsilon)&=& \bar N_3(\epsilon),
\quad X^2(\epsilon) Z_D^5(\epsilon) Z_3^2(\epsilon)= \bar
N_4(\epsilon), \quad
X^2(\epsilon) Z_D^4(\epsilon) Z_4(\epsilon)= \bar N_5(\epsilon), \nonumber\\
X(\epsilon) Z_2^2(\epsilon) Z_1^{-1}(\epsilon)Z_D(\epsilon)&=&
\bar N_6(\epsilon), \quad X(\epsilon) \tilde{Z}_2^2(\epsilon)
\tilde{Z}_1(\epsilon)Z_D(\epsilon)= \bar N_7(\epsilon),
\end{eqnarray}
where all the quantities $\bar N_i(\epsilon), \ i=1,2,3,4,5,6,7$
exist as $\epsilon$ goes to zero, so these quantities are simply
arbitrary but finite numbers, i.e., $\bar N_i(\epsilon) \equiv
\bar N_i$. Evidently, by imposing these IR convergence conditions
one requires that all the terms entering the gluon SD equation
(2.2) should survive in the $\epsilon \rightarrow 0^+$ limit,
maintaining thus its full dynamical structure. This makes it
possible not to loose even one bit of the information on the true
QCD vacuum, the dynamical and topological structures of which are
supposed to be reflected by the solutions of this equation. The
last two IR convergence conditions are known as quark and ghost
self-energy IR convergence conditions, respectively.

Let us show now that all the finite but arbitrary and different
constants $\bar N_i$ appeared in the last gluon SD equation (2.10)
as well as in the IR convergence conditions (2.11) can be put to
one not losing generality. Moreover, this is general feature of
our approach. In all the IR convergence conditions, all the
finite, but arbitrary numbers can be put to one, by simply
redefining the corresponding IRMR constants as well as the
corresponding IR renormalized quantities. In order to show this
explicitly, let us introduce indeed new IRMR constants as follows:

\begin{eqnarray}
Z_D(\epsilon) &=& \bar N_1 Z'_D(\epsilon), \quad X(\epsilon)= \bar
N_1^{-2} \bar N_2 X'(\epsilon), \nonumber\\
Z_3(\epsilon)&=& \bar N_3 \bar N_1^{-1} \bar N_2^{-1}
Z'_3(\epsilon), \quad \bar N_1^{-1} \bar N_3^2 \bar N_4^{-1} =1,
\quad
Z_4(\epsilon)= \bar N_5 \bar N_2^{-2} Z'_4(\epsilon), \nonumber\\
\bar N_1^{-1} \bar N_2 \bar N_6^{-1} Z_2^2(\epsilon)
Z_1^{-1}(\epsilon) &=& {Z'}_2^2(\epsilon) {Z'}_1^{-1}(\epsilon),
\quad \bar N_1^{-1} \bar N_2 \bar N_7^{-1} \tilde{Z}_2^2(\epsilon)
\tilde{Z}_1(\epsilon)=  \tilde{Z'}_2^2(\epsilon)
\tilde{Z'}_1(\epsilon),
\end{eqnarray}
then the IR convergence conditions (2.11) in terms of new IRMR
constants become

\begin{eqnarray}
{Z'}_D(\epsilon) &=&1, \quad X'(\epsilon) {Z'}_D^2(\epsilon)=1, \nonumber\\
X'(\epsilon) {Z'}_D^3(\epsilon) {Z'}_3(\epsilon)&=&1, \quad
{X'}^2(\epsilon) {Z'}_D^5(\epsilon) {Z'}_3^2(\epsilon)=1, \quad
{X'}^2(\epsilon) {Z'}_D^4(\epsilon) {Z'}_4(\epsilon)=1, \nonumber\\
X'(\epsilon) {Z'}_2^2(\epsilon) {Z'}_1^{-1}(\epsilon)
{Z'}_D(\epsilon)&=&1, \quad X'(\epsilon) \tilde{Z'}_2^2(\epsilon)
\tilde{Z'}_1(\epsilon)Z'_D(\epsilon)=1,
\end{eqnarray}
while the gluon SD equation (2.10) in terms of new IR renormalized
quantities is

\begin{eqnarray}
\bar D'(q) = D^0(q) &-& D^0(q)\bar T'_q(q) \bar D'(q) - D^0(q)\bar
T'_{gh}(q)\bar D'(q) + D^0(q) {1
\over 2} \bar T'_t \bar D'(q) \nonumber\\
&+& D^0(q){1 \over 2} \bar T'_1(q)\bar D'(q) + D^0(q){1 \over 2}
\bar T'_2(q) \bar D'(q) +  D^0(q){1 \over 6} \bar {T_2'}'(q) \bar D'(q). \nonumber\\
\end{eqnarray}

This simply means that all the arbitrary but finite constants
$N_i$ in the gluon SD equation (2.10) and in the IR convergence
conditions (2.12) can be put to one, i.e., $N_i=1$, indeed.
Returning then to the previous notations, the gluon SD equation
(2.10) becomes

\begin{eqnarray}
\bar D(q) = D^0(q) &-& D^0(q)\bar T_q(q) \bar D(q) - D^0(q)\bar
T_{gh}(q)\bar D(q) + D^0(q) {1
\over 2} \bar T_t \bar D(q) \nonumber\\
&+&  D^0(q){1 \over 2} \bar T_1(q)\bar D(q) +  D^0(q){1 \over 2}
\bar T_2(q) \bar D(q) + D^0(q){1 \over
6} \bar T_2'(q),  \nonumber\\
\end{eqnarray}
while the corresponding IR convergence conditions (2.11) become

\begin{eqnarray}
Z_D(\epsilon) &=&1, \quad X(\epsilon) Z_D^2(\epsilon)=1, \nonumber\\
X(\epsilon) Z_D^3(\epsilon) Z_3(\epsilon)&=&1, \quad X^2(\epsilon)
Z_D^5(\epsilon) Z_3^2(\epsilon)=1, \quad
X^2(\epsilon) Z_D^4(\epsilon) Z_4(\epsilon)=1, \nonumber\\
X(\epsilon) Z_2^2(\epsilon) Z_1^{-1}(\epsilon)Z_D(\epsilon)&=&1,
\quad X(\epsilon) \tilde{Z}_2^2(\epsilon)
\tilde{Z}_1(\epsilon)Z_D(\epsilon)=1.
\end{eqnarray}
 Evidently, the solutions of these relations are

\begin{equation}
Z_D(\epsilon) = X(\epsilon) = 1, \quad  Z_3(\epsilon) =
Z_4(\epsilon)=1, \quad Z_2^2(\epsilon) Z_1^{-1}(\epsilon)=1, \quad
\tilde{Z}_2^2(\epsilon) \tilde{Z}_1(\epsilon)=1.
\end{equation}
Thus the IRMR constants of quark and ghost degrees of freedom
remain undetermined at this stage. They will be determined
elsewhere via the corresponding ST identities, which relate them
to each other (see, for example, Refs. \cite{7,8,9,10} and
references therein). Evidently, one can start from any place in
the SD system of equations, for example, to start from the quark
and ghost sectors, ST identities, etc. However, finally the system
of the corresponding IR convergence conditions will have the same
solutions (2.17), of course.

From the solutions (2.17) and definitions (2.9) it clearly follows
that

\begin{equation}
g^2 = \bar g^2, \ D(q) = \bar D(q)  \ \Rightarrow (\xi = \bar
\xi), \ T_3(q,q_1) = \bar T_3(q,q_1), \ T_4(q,q_1,q_2) = \bar
T_4(q,q_1,q_2),
\end{equation}
i.e., all these quantities are IR renormalized from the very
beginning since all their corresponding IRMR constants are equal
to one. Thus the IR regularized (without bar) quantities in Eq.
(2.18) coincide with their IR renormalized (with bar)
counterparts. Moreover, from the fact that the full gluon
propagator is IR renormalized from the very beginning, it is easy
to understand (see Eq. (1.1)) that the gauge fixing parameter is
IR renormalized from the very beginning either (i.e., $\xi = \bar
\xi$), which is explicitly shown in Eq. (2.18). So the dependence
all of these quantities on $\epsilon$ in the $\epsilon \rightarrow
0^+$ limit can be neglected.

\section{IR renormalization of the mass gap}

In order to investigate the IR renormalization properties of the
INP part of the full gluon propagator, it is convenient to rewrite
Eq. (1.4) as follows:

\begin{equation}
D^{INP}(q, \Delta^2) =  \sum_{k=0}^{\infty} (\Delta^2)^{k+1}
(q^2)^{-2-k} \phi_k(\lambda, \nu, \xi, g^2),
\end{equation}
where we suppress the tensor $iT_{\mu\nu}(q)$, for simplicity.
Thus the IR renormalization properties of $D^{INP}(q, \Delta^2)$
depend on the mass gap $\Delta^2$ and the corresponding residues
$\phi_k(\lambda, \nu, \xi, g^2)$ only.

In our previous works \cite{1,2,3} we have derived the gluon
confinement criterion in the most general form, i.e., considering
the above-mentioned both quantities as depending on the IR
regularization parameter $\epsilon$. Our aim here is to specify
the dependence of the residues $\phi_k(\lambda, \nu, \xi, g^2)$ on
$\epsilon$ via their arguments. As a function of $\epsilon$ beside
the quantities pointed out above it may depend on the full gluon
propagator, the triple and quartic full gluon vertices (see Eqs.
(2.5)-(2.8) above), so that one has

\begin{equation}
\phi_k(\lambda, \nu, \xi, g^2) \equiv \phi_k( \lambda, \nu; g^2,
D, T_3, T_4),
\end{equation}
where instead the dependence on the gauge fixing parameter $\xi$
we show explicitly the equivalent dependence on the full gluon
propagator $D$. As mentioned above, in the presence of such severe
IR singularities (shown in Eq. (3.1)) all the quantities, which
appear in the residues, should depend in principle on $\epsilon$.
Thus, one obtains

\begin{equation}
\phi_k( \lambda, \nu; g^2, D, T_3, T_4) \equiv \phi_k(
\lambda(\epsilon), \nu(\epsilon); g^2(\epsilon), D(\epsilon),
T_3(\epsilon), T_4(\epsilon)).
\end{equation}
However, from the solutions (2.18) it follows that none of the
coupling constant squared, the full gluon propagator, the triple
and quartic full gluon vertices depend on $\epsilon$ as it goes to
zero, so we can write

\begin{equation}
\phi_k( \lambda(\epsilon), \nu(\epsilon); g^2(\epsilon),
D(\epsilon), T_3(\epsilon), T_4(\epsilon)) \equiv \phi_k(
\lambda(\epsilon), \nu(\epsilon); \bar g^2, \bar D, \bar T_3, \bar
T_4).
\end{equation}
On the other hand, the dimensionless UV cut-off $\lambda$ and the
dimensionless renormalization point $\nu$ might be functions of
the coupling constant squared $g^2$ and the gauge fixing parameter
$\xi$ since they have been introduced by hand, i.e., $\lambda =
\lambda(\xi, g^2)$ and $\nu =\nu(\xi, g^2)$. In its turn, this
means that $\lambda(\epsilon) = \lambda(\xi(\epsilon),
g^2(\epsilon))$ and $\nu(\epsilon) =\nu(\xi(\epsilon),
g^2(\epsilon))$. However, again from the solutions (2.18) it
follows that we can neglect the dependence on $\epsilon$ in the
coupling constant squared as well as in the gauge fixing
parameter. Hence the previous equation is to be present as
follows:

\begin{equation}
\phi_k( \lambda(\epsilon), \nu(\epsilon); g^2(\epsilon),
D(\epsilon), T_3(\epsilon), T_4(\epsilon)) \equiv \phi_k( \bar
\lambda, \bar \nu; \bar g^2, \bar D, \bar T_3, \bar T_4),
\end{equation}
since from $\lambda = \lambda(\xi, g^2)=  \lambda(\bar \xi, \bar
g^2) $ and $\nu =\nu(\xi, g^2)= \nu(\bar \xi, \bar g^2)$ it
follows that $\lambda = \bar \lambda$ and $\nu = \bar \nu$, where
$\bar \lambda =  \lambda(\bar \xi, \bar g^2) $ and $\bar \nu =
\nu(\bar \xi, \bar g^2)$, by definition.

That's the dimensionless UV cut-off $\lambda$ and the
dimensionless renormalization point $\nu$ are IR renormalized from
the very beginning can be in addition proven in the following way.
Let us recall that the above-mentioned quantities are the ratios
between the corresponding mass squared scale parameters and the
mass gap. As underlined above, all these squared masses depend in
general on the IR regularization parameter $\epsilon$, namely
$\Lambda^2 \equiv \Lambda^2(\epsilon)$, $\mu^2 \equiv
\mu^2(\epsilon)$ and $\Delta^2 \equiv \Delta^2(\epsilon)$. So on
general ground, one puts

\begin{eqnarray}
\Lambda^2(\epsilon) &=& \alpha(\epsilon) \Delta_1^2(\epsilon) =
\lambda \alpha_1(\epsilon) \Delta_1^2(\epsilon) = \lambda
\Delta^2(\epsilon), \nonumber\\
\mu^2(\epsilon) &=& \beta(\epsilon) \Delta_2^2(\epsilon) = \nu
\beta_2(\epsilon) \Delta_2^2(\epsilon) = \nu \Delta^2(\epsilon),
\end{eqnarray}
which assumes

\begin{equation}
\alpha_1(\epsilon) \Delta_1^2(\epsilon) = \beta_2(\epsilon)
\Delta_2^2(\epsilon) = \Delta^2(\epsilon),
\end{equation}
and which can be always satisfied, of course. Evidently, in these
relations we introduce auxiliary intermediate masses squared
$\Delta_1^2(\epsilon)$ and $\Delta_1^2(\epsilon)$. Also, the
dimensionless numbers $\lambda, \ \nu$ do not depend on
$\epsilon$, by derivation (more precisely they have finite limit
as $\epsilon$ goes to zero, by definition, i.e., $\lambda \equiv
\bar \lambda$ and $\nu \equiv \bar \nu$). Since nothing depends on
the auxiliary intermediate masses squared in our approach, all the
ratios between the initial masses squared $\Lambda^2(\epsilon), \
\mu^2(\epsilon)$ and the mass gap $\Delta^2(\epsilon)$ itself are
$\epsilon$-independent. In other words, by introducing the
above-mentioned auxiliary intermediate masses squared, we pass the
dependence on $\epsilon$ from the initial masses squared on the
mass gap, leaving the corresponding dimensionless numbers not
depending on it (evidently, this is true for any arbitrary mass
squared $M^2 \equiv M^2(\epsilon)$ by assuming the introduction of
the corresponding auxiliary intermediate mass squared). This is in
complete agreement with the above-derived result (that's none of
$\lambda$ and $\nu$ depends on $\epsilon$ in the $\epsilon
\rightarrow 0^+$ limit) which was obtained by the IR
renormalization of the gluon SD equation itself.

Going back to Eq. (3.5), it thus becomes

\begin{equation}
\phi_k( \lambda, \nu; g^2, D, T_3, T_4) \equiv \phi_k( \bar
\lambda, \bar \nu; \bar g^2, \bar D, \bar T_3, \bar T_4),
\end{equation}
since we can neglect the dependence on $\epsilon$ in all arguments
of the residues. This means that all the residues in the initial
equation (3.1) are IR renormalized from the very beginning, namely

\begin{equation}
\phi_k(\lambda, \nu, \xi, g^2) = \bar \phi_k( \bar \lambda, \bar
\nu, \bar \xi, \bar g^2).
\end{equation}
Let us recall that the corresponding IRMR constants for the
residues (see paper \cite{1}) are equal to one, i.e.,
$Z_k(\epsilon) =1$ in this case. Evidently, from here on we are
going back to the same arguments in the residues as in the
starting Eq. (3.1).

Then in Eq. (3.1) the only quantity which should be IR
renormalized remains the mass gap itself. Let us introduce further
the following relation:

\begin{equation}
\Delta^2 = X(\epsilon) \bar \Delta^2,
\end{equation}
where the mass gap with bar is IR renormalized, i.e., it exists as
$\epsilon$ goes to zero, by definition, while the mass gap without
bar is IR regularized. In complete analogy with the relations
(2.9) in both quantities the dependence on $\epsilon$ is assumed.
Here $X(\epsilon)$ is the corresponding IRMR constant.
Substituting further this relation into the Laurent expansion
(3.1), in terms of the IR renormalized quantities, it then becomes

\begin{equation}
D^{INP}(q, \bar \Delta^2) = \sum_{k=0}^{\infty} (\bar
\Delta^2)^{k+1} (q^2)^{-2-k} \bar \phi_k( \bar \lambda, \bar \nu,
\bar \xi, \bar g^2) X^{k+1}(\epsilon).
\end{equation}

\subsection{Gluon confinement}

Due to the distribution nature of severe IR singularities, which
appear in the full gluon propagator, the two different cases
should be distinguished.

{\bf I.} If there is an explicit integration over the gluon
momentum, then from the dimensional regularization (2.1) and Eq.
(3.11), it follows

\begin{equation}
D^{INP}(q, \bar \Delta^2) = \sum_{k=0}^{\infty} (\bar
\Delta^2)^{k+1} a(k)[\delta^4(q)]^{(k)} \bar \phi_k( \bar \lambda,
\bar \nu, \bar \xi, \bar g^2) \bar B_k(\epsilon),
\end{equation}
provided the INP part will not depend on $\epsilon$ at all as it
goes to zero. For this we should put

\begin{equation}
X^{k+1}(\epsilon) = \epsilon \bar B_k(\epsilon), \quad
k=0,1,2,3..., \quad \epsilon \rightarrow 0^+,
\end{equation}
then the cancellation  with respect to $\epsilon$ will be
guaranteed term by term (each NP IR singularity is completely
independent distribution) in the Laurent skeleton loop expansion
(3.11), that is dimensionally regularized and IR renormalized in
Eq. (3.12). Here $\bar B_k(\epsilon)$ exists as $\epsilon$ goes to
zero, by definition. It is easy to show that the unique solution
of the IR convergence condition (3.13) is

\begin{equation}
X(\epsilon) = \epsilon, \quad \Delta^2 = \epsilon \bar \Delta^2,
\quad  \bar B_k(\epsilon) = \epsilon^k, \quad k=0,1,2,3..., \quad
\epsilon \rightarrow 0^+,
\end{equation}
where we put $\bar B_0(\epsilon) \equiv \bar B_0 = 1$ not loosing
generality.

{\bf II.} If there is no explicit integration over the gluon
momentum, then the functions $(q^2)^{-2-k}$ in the Laurent
skeleton loops expansion (3.11) cannot be treated as the
distributions, i.e., there is no scaling as $1/ \epsilon$. The INP
part of the full gluon propagator, expressed in the IR
renormalized terms, in this case disappears as $\epsilon $, namely

\begin{equation}
D^{INP}(q, \bar \Delta^2) = \epsilon \sum_{k=0}^{\infty} (\bar
\Delta^2)^{k+1} (q^2)^{-2-k} \bar \phi_k( \bar \lambda, \bar \nu,
\bar \xi, \bar g^2) \bar B_k(\epsilon) \sim \epsilon, \quad
\epsilon \rightarrow 0^+.
\end{equation}
This means that any amplitude for any number of soft-gluon
emissions (no integration over their momenta) will vanish in the
IR limit in our picture. In other words, there are no transverse
gluons in the IR (let us remind that the INP part of the full
gluon propagator responsible for confinement of gluons within our
approach depends only on the transverse degrees of freedom of
gauge bosons), i.e., at large distances (small momenta, $q^2
\rightarrow 0$) there is no possibility to observe physical gluons
experimentally as free particles. Let us emphasize that the PT
part of the full gluon propagator is to be totally neglected in
comparison with its INP one in the $q^2 \rightarrow 0$ limit,
i.e., the full gluon propagator $D$ is reduced to its INP part in
this limit before taking the $\epsilon \rightarrow 0^+$ limit. So
color gluons can never be isolated. This behavior can be treated
as the gluon confinement criterion. It does not depend explicitly
on a gauge choice in the full gluon propagator, i.e., it is a
gauge-invariant. It is also general one, since even going beyond
the gluon sector nothing can invalidate it. For the first time it
has been derived in Ref. \cite{6} (see Ref. \cite{3} as well).
Evidently, it coincides with the general criterion of gluon
confinement derived in our previous works \cite{1,2} when the IR
renormalization properties of the residues via their arguments
have not been specified.

\section{Exact structure of the full gluon propagator}

Our quantum-dynamical approach to the true QCD ground state is
based on the existence and the importance of such kind of the NP
excitations and fluctuations of virtual gluon fields which are
mainly due to the nonlinear (NL) interactions between massless
gluons without explicitly involving some extra degrees of freedom.
They are to be summarized (accumulated) into the purely transverse
part of the full gluon propagator, and are to be effectively
correctly described by its severely singular structure in the deep
IR domain. We will call them the purely transverse singular gluon
fields, for simplicity. In other words, they represent the purely
transverse quantum virtual fields with the enhanced low-frequency
components/lagre scale amplitudes due to the NL dynamics of the
massless gluon modes.

At this stage, it is difficult to identify actually which type of
gauge field configurations can be finally formed by the purely
transverse singular gluon fields in the QCD ground state, i.e., to
identify relevant field configurations: chromomagnetic, self-dual,
stochastic, etc. However, if these gauge field configurations can
be absorbed into the gluon propagator (i.e., if they can be
considered as solutions to the corresponding SD equation), then
its severe IR singular behavior is a common feature for all of
them. Being thus a general phenomenon, the existence and the
importance of quantum excitations and fluctuations of severely
singular IR degrees of freedom inevitably lead to the general zero
momentum modes enhancement (ZMME) effect in the QCD ground state.

Our approach to the true QCD ground state, based on the general
ZMME phenomenon there, can be thus analytically formulated in
terms of the exact decomposition of the full gluon propagator. In
order to define correctly the NP phase in comparison with the PT
one in QCD, let us introduce (following our paper \cite{3}) the
exact decomposition of the full gluon form factor (which in
principle can be treated as the effective charge) as follows:

\begin{equation}
d(q^2, \xi) = d(q^2, \xi) - d^{PT}(q^2, \xi) + d^{PT}(q^2, \xi) =
d^{NP}(q^2, \xi) + d^{PT}(q^2, \xi).
\end{equation}
Evidently, $d(q^2, \xi)$ being the NP effective charge,
nevertheless, is contaminated by the PT contributions, while
$d^{NP}(q^2, \xi)$ is the truly NP one since it is free of them,
by construction. Substituting now this decomposition into the full
gluon propagator (1.1), one obtains

\begin{equation}
D_{\mu\nu}(q) = D^{INP}_{\mu\nu}(q)+ D^{PT}_{\mu\nu}(q),
\end{equation}
where

\begin{equation}
D^{INP}_{\mu\nu}(q) = i T_{\mu\nu}(q) d^{NP}(q^2, \xi) {1 \over
q^2} =i T_{\mu\nu}(q) d^{INP}(q^2, \xi),
\end{equation}
and

\begin{equation}
D^{PT}_{\mu\nu}(q) = i \Bigr[ T_{\mu\nu}(q) d^{PT}(q^2, \xi)+ \xi
L_{\mu\nu}(q) \Bigl] {1 \over q^2}.
\end{equation}
As mentioned above, the PT part, denoted in Eq. (1.3) as
$O_{\mu\nu}(1/q^2)$, remains undetermined. At the same time, the
INP part (representing the above-mentioned ZMME effect) in terms
of the IR renormalized quantities is

\begin{equation}
D^{INP}_{\mu\nu}(q, \bar \Delta^2) = i T_{\mu\nu}(q) \times
\epsilon \sum_{k=0}^{\infty} (\bar \Delta^2)^{k+1} (q^2)^{-2-k}
\bar \phi_k( \bar \lambda, \bar \nu, \bar \xi, \bar g^2) \bar
B_k(\epsilon).
\end{equation}
However, it is perfectly clear now that due to solutions (3.14)
only the simplest NP IR singularity $(q^2)^{-2}$ will survive in
the $\epsilon \rightarrow 0^+$ limit, namely

\begin{equation}
D^{INP}_{\mu\nu}(q, \bar \Delta^2) = i T_{\mu\nu}(q) \times
\epsilon \bar \Delta^2 (q^2)^{-2}\bar \phi_0( \bar \lambda, \bar
\nu, \bar \xi, \bar g^2),
\end{equation}
since all other terms in the expansion (4.5) become terms of the
order $\epsilon^2$, at least, in the $\epsilon \rightarrow 0^+$
limit. Here

\begin{equation}
\bar \phi_0( \bar \lambda, \bar \nu, \bar \xi, \bar g^2) =
\sum_{m=0}^{\infty} \bar \phi_{0,m}( \bar \lambda, \bar \nu, \bar
\xi, \bar g^2).
\end{equation}
Thus an infinite number of iterations of the relevant skeleton
loops gives rise to a simplest NP IR singularity (and hence to the
mass gap).

{\bf I.} Again, if there is an explicit integration over the gluon
momentum, then on account of the regularization relation (2.1) for
$k=0$ and $a(0) = \pi^2$, one finally gets

\begin{equation}
D^{INP}_{\mu\nu}(q, \bar \Delta^2)= i T_{\mu\nu}(q) \times \bar
\Delta^2_R \delta^4(q).
\end{equation}
The $\delta$-type regularization of the simplest NP IR singularity
$(q^2)^{-2}$ is valid even for the multi-loop skeleton diagrams,
where the number of independent loops is equal to the number of
the gluon propagators. In the multi-loop skeleton diagrams, where
these numbers do not coincide (for example, in the diagrams
containing three or four-gluon proper vertices), the general
regularization (2.1) is to be used, i.e., the derivatives of the
$\delta$ functions, which should be understood in the sense of the
theory of distributions \cite{5} (for detail prescription how to
correctly proceed in this case see our paper \cite{6}). In Eq.
(4.8) we introduce the UV renormalized (which has been already the
IR renormalized) mass gap as follows:

\begin{equation}
\bar \Delta^2_R = Z_{\bar \Delta}(\bar \lambda, \bar \nu, \bar
\xi, \bar g^2) \bar \Delta^2 (\bar \lambda, \bar \nu, \bar \xi,
\bar g^2),
\end{equation}
and

\begin{equation}
Z_{\bar \Delta}(\bar \lambda, \bar \nu, \bar \xi, \bar g^2) =
\Bigl[ \pi^2 \sum_{m=0}^{\infty} \bar \phi_{0,m}(\bar \lambda,
\bar \nu, \bar \xi, \bar g^2) \Bigr],
\end{equation}
where $Z_{\bar \Delta}(\bar \lambda, \bar \nu, \bar \xi, \bar
g^2)$ is the corresponding UVMR constant, and $\bar \Delta^2_R$
exists in the $\bar \lambda \rightarrow \infty$ limit, by
definition. It is also a gauge-invariant since nothing in its
definition depends explicitly on the gauge fixing parameter.
Precisely this quantity should be considered as the physical mass
gap, which must be strictly positive, by definition. Let us
emphasize that it survives after summing up an infinite number of
the relevant contributions (skeleton loops expansion) and
performing the IR renormalization program. This is similar to
$\Lambda^2_{QCD}$, which also appears after summing up an infinite
number of the relevant contributions by solving the
renormalization group equations for the effective coupling in the
weak coupling regime and taking the $\lambda \rightarrow \infty$
limit. At the same time, the dependence of the mass gap on the
coupling constant squared $g^2$ is completely arbitrary because of
an infinite summation of the relevant skeleton loop integrals, so
its all orders contribute into the mass gap. In other words, it
plays no role in the presence of the mass gap. Let us also
emphasize that the UVMR constant $Z_{\bar \Delta}(\bar \lambda,
\bar \nu, \bar \xi, \bar g^2)$, introduced in Eq. (4.9) and
defined in Eq. (4.10), itself is an infinite sum over all NL
iterations of the relevant dimensionless skeleton loop integrals,
and hence it cannot be calculated perturbatively. It is
essentially NP UVMR constant, by its nature.

{\bf II.} Again, if there is no explicit integration over the
gluon momentum, then the criterion of gluon confinement (3.15)
remains valid, of course. Now it looks like

\begin{equation}
D^{INP}_{\mu\nu}(q, \bar \Delta^2) = \epsilon \times i
T_{\mu\nu}(q) \bar \Delta^2 (q^2)^{-2} \bar \phi_0( \bar \lambda,
\bar \nu, \bar \xi, \bar g^2)  \sim \epsilon, \quad \epsilon
\rightarrow 0^+,
\end{equation}
in complete agreement with Eq. (4.6). Let us emphasize once more
that it takes place at any gauge, and thus is gauge-invariant.

So, we have established the exact structure of the INP part of the
full gluon propagator which is responsible for color confinement
of gluons within our approach. In all loop integrals for the
independent loop variable it is explicitly given in Eq. (4.8). In
all other cases the derivatives of the $\delta$ function are in
order as mentioned above. All this has been achieved at the
expense of the PT part of the full gluon propagator which remains
undetermined. However, this is not important within our approach
(see discussion below in section V).

The ZMME mechanism of quark confinement is nothing but the well
forgotten IR slavery (IRS) one, which can be equivalently referred
to as a strong coupling regime \cite{7,11}. Indeed, at the very
beginning of QCD it was expressed a general idea
\cite{11,12,13,14,15,16} that the quantum excitations of the IR
degrees of freedom, because of self-interaction of massless gluons
in the QCD vacuum, made it only possible to understand
confinement, dynamical (spontaneous) breakdown of chiral symmetry
and other NP effects. In other words, the importance of the deep
IR structure of the true QCD vacuum has been emphasized as well as
its relevance to the above-mentioned NP effects and the other way
around. This development was stopped by the wide-spread wrong
opinion that severe IR singularities cannot be put under control.
We have explicitly shown (see our recent papers \cite{1,2,3,6} and
references therein) that the correct mathematical theory of
quantum YM physical theory is the theory of distributions (the
theory of generalized functions) \cite{5}, complemented by the DR
method \cite{4}. They provide a correct treatment of these severe
IR singularities without any problems. Thus, we come back to the
old idea but on a new basis that is why it becomes new ("new is
well forgotten old"). In other words, we put the IRS mechanism of
quark confinement on a firm mathematical ground provided by the
distribution theory. Moreover, we also emphasize the role of the
purely transverse singular gauge fields in this mechanism.

Working always in the momentum space, we are speaking about the
purely transverse singular gluon fields responsible for color
confinement in our approach. Discussing the relevant field
configurations, we always will mean the functional (configuration)
space. Speaking about relevant field configurations
(chromomagnetic, self-dual, stochastic, etc), we mean all the
low-frequency modes of these virtual transverse fields. Only large
scale amplitudes of these fields ("large transverse gluon fields")
are to be taken into account by the INP part of the full gluon
propagators. All other frequencies are to be taken into account by
corresponding PT part of the gluon propagators. Apparently, to
speak about specific field configurations that are solely
responsible for color confinement is not the case, indeed. The
low-frequency components/large scale amplitudes of all the
possible in the QCD vacuum the purely transverse virtual fields
are important for the dynamical and topological formation of such
gluon field configurations which are responsible for color
confinement and other NP effects within our approach to low-energy
QCD. For convenience, we will call them the purely transverse
severely singular gluon field configurations as mentioned above.

\subsection{A few technical remarks}

The exact separation of the full gluon propagator into the two
principally different parts, shown in Eq. (4.2), does not, of
course, contradict to the full gluon propagator being IR
renormalized from the very beginning ($D = \bar D$ and hence
$Z_D(\epsilon)=1$). If there is no explicit integration over the
gluon momentum $q$, then $D$ is reduced to $D^{PT}$ which implies
$D =D^{PT} = \bar D^{PT}$, indeed. On the other hand, if there is
an explicit integration over the gluon momentum $q$, then,
nevertheless, the INP part has a finite limit as $\epsilon$ goes
to zero (see Eq. (3.12)). So $D=\bar D$ will be again satisfied.
Also, there is no doubt that our solution for the full gluon
propagator, obtained at the expense of remaining unknown its PT
part, nevertheless, satisfies the gluon SD equation (2.15) apart
from the quark and ghost skeleton loops since it has been obtained
by the direct iteration solution of this equation. To show this
explicitly by substituting it back into the gluon SD equation
(2.15) is not a simple task, and this is to be done elsewhere (for
preliminary procedure see our paper \cite{6}). The problem is that
the decomposition of the full gluon propagator into the INP and PT
parts by regrouping the so-called mixed up terms in Ref. \cite{1}
(see also Refs. \cite{3,6}) was a well defined procedure (there
was an exact criterion how to distinguish between these two terms
in a single $D$). However, to do the same at the level of the
gluon SD equation itself, which is nonlinear in $D$, is not so
obvious.

Fortunately, there exists a rather simple method as how to show
explicitly that the INP part of the full gluon propagator can be
completely decoupled from the rest of the gluon SD equation in the
$\epsilon \rightarrow 0^+$ limit. In other words, let us consider
it as a function of $\epsilon$ rather than as a function of its
momentum. For this purpose, let us present explicitly the gluon SD
equation which was the starting point for the general iteration
solution for the full gluon propagator in Ref. \cite{1}, namely

\begin{equation}
D(q) = D^0(q) + D^0(q)T_g[D]D(q) + D^0(q)O(q^2;D)D(q).
\end{equation}
From its iteration solution we already know that on general ground
the block $T_g[D]$ can be represented as follows:

\begin{equation}
T_g[D] = \Delta^2 \sum_{k=0}^{\infty} \sum_{m=0}^{\infty}
b_{k,m}(\lambda, \nu, \xi, g^2),
\end{equation}
where the $q^2$-independent factors $b_{k,m}(\lambda, \nu, \xi,
g^2)$ may only depend on the same arguments as the real residues
$\phi_{k,m}(\lambda, \nu, \xi, g^2)$ in Eq. (3.2). From the IRMR
program formulated and developed here (see sections II and III) it
follows that just as the real residues the $q^2$-independent
factors do not depend on $\epsilon$ in the $\epsilon \rightarrow
0^+$ limit. Taking this into account, and going to the IR
renormalized quantities in Eq. (4.13), one obtains

\begin{equation}
T_g[\bar D] = \epsilon \bar \Delta^2 \sum_{k=0}^{\infty}
\sum_{m=0}^{\infty} \bar b_{k,m}(\bar \lambda, \bar \nu, \bar \xi,
\bar g^2) \sim \epsilon, \quad   \epsilon \rightarrow 0^+,
\end{equation}
where, evidently, we equivalently replace all quantities with
their IR renormalized counterparts since $D = \bar D$ and so on
(let us emphasize that after doing so here and everywhere, only
then we can go to the $\epsilon \rightarrow 0^+$ limit). In other
words, this block disappears as $\epsilon$ in the $\epsilon
\rightarrow 0^+$ limit (at any $k$), so Eq. (4.12) becomes

\begin{equation}
\bar D(q) = D^0(q) + D^0(q)O(q^2;\bar D) \bar D(q).
\end{equation}
Now we can distinguish between the INP and PT parts of the full
gluon propagator shown in Eq. (4.2). Since there is no explicit
integration over the gluon momentum $q$, the INP part of the full
gluon propagator $\bar D$ vanishes as $\epsilon$ in the $\epsilon
\rightarrow 0^+$ limit  (see Eq. (4.11)). So from Eq. (4.15) it
finally follows

\begin{equation}
\bar D^{PT}(q) = D^0(q) + D^0(q)O(q^2; \bar D) \bar D^{PT}(q).
\end{equation}
Let us underline that this equation will produce only the PT-type
of the IR singularities within its iteration solution since the
block $O(q^2; \bar D)$ is always of the order $q^2$ whatever $\bar
D$ is. Thus the INP part of the full gluon propagator
automatically satisfies the gluon SD equation. The part of the
gluon SD equation responsible for the NP IR singularities in its
general iteration solution vanishes on account of the solution for
its INP part. Just in this sense should be understood in general
terms the solution of the gluon SD equation within our approach
since it leaves the PT part of the solution undetermined. To show
this explicitly by treating the INP part of the full gluon
propagator as a function of its momentum and substituting the
decomposition (4.2) back to the nonlinear gluon SD equation (4.12)
is far more complicate case as mentioned above.

\hspace{3mm}

For the calculations of such NP quantities as the gluon condensate
or the truly NP vacuum energy density (the Bag constant apart from
the sign, by definition), the effective coupling constant
(effective charge) should be used from the very beginning. From
the fact that within our approach the simplest NP IR singularity
$(q^2)^{-2}$ survives only and in our notations (see Eq. (4.3)) it
then follows

\begin{equation}
\alpha_s(q^2) = q^2d^{INP}(q^2) = \Lambda^2_{NP}/ q^2,
\end{equation}
where $\Lambda^2_{NP}$ is identified with $\bar \Delta^2_R$, for
simplicity (a possible difference between them is not important).
The fact that the effective charge (4.17) has a PT IR singularity
from the very beginning makes it formally possible to put
$\Delta^2 = \bar \Delta^2$ (up to some as mentioned above
unimportant finite constant), i.e., to omit the dependence on
$\epsilon$ in Eq. (4.6). Though the gluon condensate is not
directly measured quantity, it enters many physical relations,
that is why it should depend on $\Lambda^2_{NP}$. For its correct
calculations (free from the PT "contaminations") see, for example,
papers \cite{17,18}. The renormalization group equation which
determines the corresponding $\beta$ function and its solution for
this effective charge is

\begin{equation}
q^2 { d \alpha_s(q^2) \over dq^2} = \beta(\alpha_s(q^2)) = -
\alpha_s(q^2),
\end{equation}
so that the $\beta$ function as a function of the effective charge
is always in the domain of attraction (i.e., negative) as it is
required for the confining theory \cite{7}. Also, in order to
calculate the linear rising potential between heavy quarks the INP
gluon form factor shown in Eq. (4.17) is to be used.

\section{Discussion and conclusions}

A few years ago Jaffe and Witten (JW) have formulated the
following theorem \cite{19}:

\vspace{3mm}

 {\bf Yang-Mills Existence And Mass Gap:} Prove that
for any compact simple gauge group $G$, quantum Yang-Mills on
$\bf{R}^4$ exists and has a mass gap $\Delta > 0$.

\vspace{3mm}

Of course, at present to prove the existence of the YM theory with
compact simple gauge group $G$ is a formidable task yet. It is
rather mathematical than physical problem. However, the general
result of our investigation in Refs. \cite{1,2,3} and here can be
formulated similar to the above-mentioned JW theorem as follows:

\vspace{3mm}

{\bf Yang-Mills Existence, Mass Gap And Gluon Confinement:} If
quantum Yang-Mills with compact simple gauge group $G=SU(3)$
exists on $\bf{R}^4$, then it exhibits a mass gap and confines
gluons.

\vspace{3mm}

Though our mass gap (4.9) reproduces many features of the JW mass
gap \cite{19}, nevertheless, the latter is more general
conception, at least at this stage (see below). The symbolic
relation between our mass gap ($\bar \Delta_R \equiv
\Lambda_{NP}$), the JW one ($\Delta \equiv \Delta_{JW}$) and
$\Lambda_{QCD} \equiv \Lambda_{PT}$ is

\begin{equation}
\Lambda_{NP} \longleftarrow^{\infty \leftarrow \alpha_s}_{0
\leftarrow M_{IR}} \ \Delta_{JW} \ { }^{\alpha_s \rightarrow
0}_{M_{UV} \rightarrow \infty} \longrightarrow  \ \Lambda_{PT},
\end{equation}
where $\alpha_s$ is obviously the fine structure coupling constant
of strong interactions, while $M_{UV}$ and $M_{IR}$ are the UV and
IR cut-offs, respectively. The right-hand-side limit is well known
as the weak coupling regime, while the left-hand-side can be
regarded as the strong coupling regime. We know how to take the
former \cite{7,11}, and we hope that we have explained in Refs.
\cite{1,2,3} and here how to deal with the latter one, not solving
the gluon SD equation directly, which is formidable task, anyway.
However, there is no doubt that the final goal of this limit,
namely, the mass gap $\Lambda_{NP}$ exists, and should be the
renormalization group invariant in the same way as
$\Lambda_{QCD}$. It is solely responsible for the large scale
structure of the true QCD ground state. It is important to
emphasize once more that it has not been introduced by hand. We
have explicitly shown that it was hidden in the skeleton loop
integrals contributing into the gluon self-energy due to the NL
interaction of massless gluon modes (Eqs. (2.5)-(2.8)). The mass
gap shows explicitly up when the gluon momentum goes to zero. The
appropriate regularization procedure has been applied to make the
existence of the mass gap perfectly clear. Moreover, it survives
an infinite summation of the corresponding skeleton loop
contributions (skeleton loop expansion) after completing the
general IRMR program.

Let us continue our discussion recalling that many important
quantities in QCD, such as the gluon and quark condensates, the
topological susceptibility, the Bag constant, etc., are defined
beyond the PT only \cite{20,21,22}. This means that they are
determined by such $S$-matrix elements (correlation functions)
from which all types of the PT contributions should be, by
definition, subtracted. Anyway, our theory for low-energy QCD
which we call INP QCD \cite{6} will be precisely defined by the
subtraction of all types of the PT contributions. At the
fundamental (microscopic) gluon propagator level the first
subtraction is provided by the exact decomposition (4.1). The
second one is to omit the PT part of the full gluon propagator
$D^{PT}$. Then one obtains the full gluon propagator free from all
types of the PT "contamination" (that is why the PT part is not
important within our approach as underlined above). Only after
these subtractions one can identify our mass gap $\Lambda_{NP}$
with the JW mass gap $\Delta_{JW}$ (for more detailed discussion
of the necessary subtractions see Ref. \cite{6}). At the same
time, having made these subtractions, we thus know the full gluon
propagator (4.8) and its generalizations (the derivatives of the
$\delta$ function) which can be used for the solutions of the
quark SD equation, quark-gluon ST identity, etc., \cite{6,7,8}.
This opens the possibilities to calculate physical observables
from first principles (for preliminary calculations see our papers
\cite{23,24}).

 Color confinement of gluons is the IR renormalization of the mass
gap gauge-invariant effect within our approach. An infinite number
of iterations of the relevant skeleton loops (skeleton loops
expansion) has to be made in order to invoke the mass gap. No any
truncations/approximations have been made as well in such obtained
general iteration solution of the gluon SD equation for the full
gluon propagator. The important feature of our investigation,
that's, the existence of the mass gap assumes certainly
confinement of gluons, somehow has been missed from the discussion
in Ref. \cite{19}. However, it is worth emphasizing that our mass
gap and the JW mass gap cannot be interpreted as the gluon mass,
i.e., they always remain massless. Our gluon propagator, described
in the previous section IV, takes into account the importance of
the quantum excitations of severely singular IR degrees of freedom
in the true QCD vacuum. They lead to the formation of the purely
transverse severely singular gluon field configurations there.
Just these configurations are primary responsible for the NP
effects, such as color confinement, dynamical chiral symmetry
breaking, etc., within our approach.

As mentioned above, we have explicitly shown that in the initial
Laurent loops expansion (3.1), that is dimensionally regularized
in Eq. (3.12), the simplest severe IR singularity $(q^2)^{-2}$
survives only (see Eq. (4.6)). So, we have confirmed and thus
revitalized the previous investigations \cite{9,10,25,26,27,28,29}
(and references therein), in which this behavior has been obtained
as asymptotic solution to the gluon SD equation in different
gauges. Let us emphasize, however, that our result is exact. We
have already shown that it leads to quark confinement and
spontaneous (dynamical) breakdown of chiral symmetry \cite{22,29}
(and references therein). In this connection let us note in
advance that quark and ghost degrees of freedom play no any
significant role in the dynamical generation of the mass gap in
our approach. As explicitly shown here and in our previous works
\cite{1,2,3}, the NL interaction of massless gluon modes is only
important. At the same time, the quark and ghost skeleton loop
contributions into the gluon self-energy (see Eqs. (2.3) and
(2.4)) do not depend on the full gluon propagator. As a result
their contributions can be summed up into the geometrical
progression series within the corresponding linear iteration
procedure. All this will complicate the IRMR program from a
technical point of view only, and it is left to be done elsewhere
(see also Ref. \cite{6}).

The smooth in the IR gluon propagator is also possible depending
on different truncations/approximations used \cite{30} (see papers
\cite{31,32} and references therein as well) since the gluon SD
equation is highly nonlinear one. The number of solutions for such
kind of systems is not fixed $a \ priori$. The singular and smooth
in the IR solutions for the gluon propagator are independent from
each other, and thus should be considered on equal footing.
However, the smooth gluon propagator is rather difficult to relate
to color confinement in a gauge-invariant way, in particular to
gluon confinement, while severely IR singular one is directly
related to it as explicitly demonstrated here and in our previous
works \cite{1,2,3,6} (and references therein).

It is interesting to note that Gribov \cite{33} by differentiating
twice the quark SD equation in fact arrived at the same
$\delta$-type potential (4.8) as well. However, the principal
feature of our approach -- the mass gap -- was missing in this
procedure.

Support from HAS-JINR Scientific Collaboration Fund (P. Levai) is
to be acknowledged. I would like to thank P. Forgacs for bringing
my attention to the Jaffe-Witten presentation of the Millennium
Prize Problem in Ref. \cite{19}. It is a great pleasure also to
thank A. Jaffe for bringing my attention to the revised version of
the above-mentioned presentation.

\end{document}